\documentclass[fleqn,10pt]{wlscirep}
\usepackage[utf8]{inputenc}
\usepackage[T1]{fontenc}
\usepackage{stmaryrd} 
\newcommand{\xn}{\noindent}
\newcommand{\xv}{\vspace{0.1cm}}
\newcommand{\br}{\text{\tiny BR}}
\newcommand{\mr}{\text{\tiny MR}}
\newcommand{\stdev}{standard deviation~}

\newcommand{\var}{variance~}
\newcommand{\fkt}{\frac{k}{2}}
\title{Natural Averaging May Complement  Known Biological Constraints in Bi-parental Reproduction’s Advantages Over Mono-parental in \\ Conserving Species Quantitative Traits}

\author[1,*]{Assaf Marron}
\author[1]{Smadar Szekely}
\author[2]{Irun R. Cohen}
\author[1]{David Harel}
\affil[1]{Department of Computer Science and Applied Mathematic, Weizmann Institute of Science, Rehovot, 76100, Israel}
\affil[2]{Department of  Immunology and Regenerative Biology,  
Weizmann Institute of Science, Rehovot, 76100, Israel}
\affil[*]{assaf.marron@weizmann.ac.il}


\begin{abstract}
Commonly recognized evolutionarily relevant effects of sexual reproduction include increased diversity, accelerated adaptation, and constrained accumulation of deleterious mutations, along with a secondary effect of species genotype homogenization. 
Still, strong published arguments prioritize the contribution of biological mechanisms underlying bi-parental reproduction to maintaining species identity above their contribution to diversity. Here, we contribute to the latter position.
In an initial mathematical analysis and simulation, we show that in an environment where copying is prone to error, quantitative polygenic traits that are shared within a parents' generation are transmitted to future generations under bi-parental reproduction with less deviation than under asexual reproduction.
Furthermore, we abstract away many biological details, and show that this trait conservation is a general statistical effect, driven by the very nature of mixing of parental traits, separately from DNA repair and from the reproductive failures, barriers and disadvantages induced by biological mechanisms.  Since survival of ecosystem interaction networks depends on the ability of individuals to  replace the networked function of failing, dying or absent members of the same species, more faithful inheritance of common traits helps sustain species and ecosystems. This sustaining effect may have contributed to the very evolution of sexual reproduction.
\end{abstract}
\begin{document}

\flushbottom
\maketitle
%
%
\thispagestyle{empty}
\section{Introduction}\label{sec:introduction}

There are multiple theories for the evolution of bi-parental reproduction, both sexual (with two or more mating types) 
and unisexual (with one mating type, as is common, for example in fungi~\cite{Heitman2015Evolution}), 
as contrasted with mono-parental, clonal, asexual reproduction. However, the underlying question is still considered open~\cite{macpherson2023sexualReproParadox,macpherson2021sexOpenProblem,livnatPapaDimitriou2016SexAsAnAlgorithm,billiard2012sexUnsolvedQuestions}.
Presently, in scientific publications and popular discussions covering the drivers for evolution of sexual reproduction, or the consequences of  various types of  bi-parental reproduction, the most commonly listed aspects are: 
\begin{enumerate}
\item  Introduction of diversity, which then leads to increased adaptivity to changing conditions, enhanced competitiveness and the ability to co-evolve with pathogens, parasites, or symbionts; for example, the Red Queen hypothesis~\cite{vanValen1973Redqueen};  

\item Constraining of the irreversible penetration of deleterious mutations, often termed Muller's ratchet~\cite{muller1932ratchettGeneticAspectsOfSex}; 
this occurs by combination of dominant/recessive relations of pairs of alleles, and by the reproductive disadvantage imposed by the deleterious trait at hand, which may be augmented by  Kondrashov's Hatchet\cite{kondrashov1988deleterious} -- interactions between slightly deleterious mutations; and, 

\item Preservation of common traits through biological processes including (i)~the raising of reproduction barriers between divergent subgroups within a species~\cite{bartonBrigsgEtAl2007evolutionBook},
(ii)~DNA repair that occurs as part of meiosis~\cite{bernstein1981repair,bartonBrigsgEtAl2007evolutionBook},
(iii)~sexual selection which promotes reproduction of individuals with certain preferred traits\cite{bartonBrigsgEtAl2007evolutionBook}, and (iv)~interference with specialization and speciation of variants that find and prefer new niches~\cite{bartonBrigsgEtAl2007evolutionBook}(ch.16).
\end{enumerate}
The relative importance or prominence of the various factors is under debate. For example, in~\cite{HengGorelick2011sexAndSpeciesIdentity}, Heng and Gorelick first observe that 
diversity, adaptivity, and constraining of deleterious mutations are generally considered the more prominent factors, while the contribution of sexual reproduction to faithful copying and genotype homogenization is relegated to a secondary or supplementary role; they then argue that much greater significance should be assigned to the latter, as these processes maintain the species identity -- the functions that constitute the delicate interaction networks that sustain the species and its ecosystem. 
This view can be aligned with the observation in~\cite{cohen2016updatingDarwin} that one of the effects of sexual reproduction is \emph{``to maintain a frequency distribution of genes within the species that fits the life style and ecosystem''}.

In this paper we contribute to this emphasis on 
retention of common parent-generation traits (hereafter abbreviated as \emph{Retention of Common Traits} (RoCT))
with generalized statistical analysis and simulation. 
We show that in a population where any kind of copying is imperfect, bi-parental reproduction that includes merging of encodings of inherited traits, conserves in the offspring generation traits that are common within the parents' generation better than does mono-parental reproduction. Furthermore, this effect is a general statistical property, driven by the very nature of mixing of parental traits, and distinct from the biological processes of repair and limits on reproduction. 
This, of course, then extends to conservation of traits across multiple generations.
Given that survival of organisms, species and ecosystems depends on interaction networks, and that all organisms eventually die, higher fidelity RoCT helps sustain species by enabling offspring to step in and carry out interaction roles performed by their ancestors.

While mathematical/statistical comparisons of RoCT under mono- and bi-parental reproduction, separately from the particular biological mechanisms, are quite sparse in recent literature, certain aspects of the topic were discussed as far back as a century ago.  
For example, in 1918, Fisher~\cite{fisher1918quantitative} establishes the correlation between the phenotypic expression of quantitative traits (also termed \emph{characters}) in offspring and that of the parents.  In 1886, Galton~\cite{galton1886regressionTowardsMediocrity} reports observations that for quantitative traits, offspring traits tended towards the mean of the population, eliminating extremes that may have been manifested by their parents, a phenomenon termed ``regression towards the mean''.  
In his experiment measuring human stature he also observes population-wide processes which he summarizes as \emph{``two opposite sets of actions, one concentrative and the other dispersive"}, but concludes that \emph{``they necessarily neutralise one another, and fall into a state of stable equilibrium''}\cite{galton1886regressionTowardsMediocrity}(p.256). 

Hamilton~\cite{hamilton1980sexVsNoSexVsParasite}, observes that 
\emph{``sex reduces the fitness
variance in frequency-dependent host-parasite systems''}; this may appear relevant to the present discussion if fitness is equated with the trait at hand. However, this model is unique in that the fitness variance is reduced in an environment where a parasite evolves to attack the phenotype with highest frequency, because sexual reproduction creates a population with diverse traits, enabling evasion of the parasite's attacks. 

Doebeli~\cite{doebeli1996quantitativeGeneticsPopulationDynamics} compares mathematical models for dynamics of population size, i.e., density fluctuations, under various parameters including population diversity, and under sexual and asexual reproduction. The focus is on quantitative phenotypic traits, which are manifested as fitness. The entire setting is that of competition and natural selection, with models addressing competition between individuals in the population and between sexuals and asexuals within the population. The model also considers population regulation through host-parasite interactions. The models and simulations show that sexual reproduction reduces density fluctuations and fitness variance within a generation. One may also infer that the fitness of an offspring generation is closer to that of the parent generation under sexual reproduction than under asexual reproduction, which alludes in some ways to RoCT.

Such analyses focus either on statistical analysis of trait inheritance in the absence of mutation, or on the evolution of traits when both genetic changes and natural selection, or fitness, play a central role. The present paper presents an analytical and simulation perspective on  inherent advantages of sexual reproduction over asexual in preserving common traits when the copying mechanisms are imperfect, say, due to mutations, but in the absence of biological considerations, i.e., not considering repair mechanisms or whether the trait at hand contributes to or interferes with reproductive success. 

RoCT may be correlated with homogenization  of offspring genotype, but these are distinct concepts. Indeed, the fact that bi-parental reproduction results in homogenization of the genome in the offspring generation is often taken for granted when a population is panmictic, i.e., every organism of the species can mate with every other of the same species, and all are confined within a geographical region (see, e.g., the abstract of~\cite{fitzpatrick2008SympatricAbstract}); it is also often implicit from discussions of gene flow in structured populations (see, e.g., ch.16, p.441 in~\cite{bartonBrigsgEtAl2007evolutionBook}), or in pondering the puzzle of sympatric speciation: why panmictic populations sometimes give rise to diversity and speciation even within a given location~\cite{fitzpatrick2008SympatricAbstract,mayr1982biologicalThought}. However, mere homogenization within the offspring generation alone does not immediately imply retention of parental traits and may even dilute such traits. 

As in the research by Doebeli~\cite{doebeli1996quantitativeGeneticsPopulationDynamics}, our analysis and simulation is constrained to quantitative traits, which by and large are polygenic. 
Quantitative traits constitute the majority of the identifiable traits in organisms~\cite{neal2018populationbiology}(p.239), and our knowledge about how various genes affect each such phenotypic trait is very limited. Indeed, the very existence of quantitative phenotypic traits whose  manifestation can take into account summing or averaging the traits of the two parents is an important factor that enables the retention of what is common or shared in the entire generation of these parents. 

The paper is structured as follows: In Section~\ref{sec:ModelingRationale}, we discuss some of the motivation for the particular principles applied in the ensuing modeling, analysis and simulation examples.  
In Section~\ref{sec:math} we  introduce a model for genotype, phenotype and sexual and asexual transmittal of quantitative characters. In Section~\ref{sec:comparison} we prove, using basic probability and the central limit theorem in a restricted synthetic model, the inevitability of conservation of quantitative parental traits  as a general effect of bi-parental reproduction, that is independent of biological specifics, and is more general than known genotype homogenization mechanisms. 
In Section~\ref{sec:simulations} we illustrate the differences between the RoCT effect under bi- and mono-parental reproduction with examples drawn from simulation runs that reproduce diverse objects including abstract sets of points, images of printed text, and images of flowers. The latter simulations also illustrate how conservation of many distinct traits, like individual Red-Green-Blue (RGB) values in thousands of pixels, can affect conservation of emergent traits, like recognizing that an image is of a flower. 

Further studies are needed to generalize and expand the model, and to confirm that the improved RoCT afforded by bi-parental reproduction is indeed a distinct natural phenomenon.

\section{Modeling Rationale}\label{sec:ModelingRationale}

Before delving into the details of the mathematical model and the illustrating simulations, in this section we offer some background and context for various choices applied in the model and its analysis. The reader may choose to skip this section, and return to it as needed. 

We support our claim that the effects of bi-parental reproduction are intrinsic, by initially showing that they exist in an abstracted setting: we deal with only a single quantitative trait; we do not attempt to model in detail transmittal of non-quantitative traits, or the processes of meiosis, genetic recombination, epistasis, and syngamy, the role of mating types, the role of the environment, or other complex aspects of reproduction in nature.
While all these factors are important and should be part of realistic models of reproduction in nature, 
the particular aspects of bi-parental reproduction discussed here deserve a special focus beyond the published theories, mechanisms, and observed effects. 

\subsection{Codes, interaction codes, and autoencoding}

The present work is part of a larger research project. In~\cite{cohenMarron2023autoencoding,cohenMarron2020survivalOfTheFitted} and related papers we have outlined how sustainment of a species depends on interaction networks in which it participates. In such networks, the role of each species is naturally and automatically encoded in what we term \emph{species interaction code}, manifested in the genes, physiology and environment of the individuals of the species. As interactions depend on traits that enable them, faithful transmittal is key.

An important assumption in the analysis is that to be transmitted, a phenotypic trait, like height,  life span, or coloring, must first be encoded. The encoding itself does not have to be compact or minimalistic; for example, it may involve extensive redundancy in order to 
overcome decoding errors, or it may be captured physically, as is the case in a human-made diagram, or graph,  of a simple relationship between a few objects, which may be physically recorded in millions of pixels.  

This perspective connects the present work to quantitative genetics~\cite{bartonBrigsgEtAl2007evolutionBook} $^{(ch.14,16,17,19,26),}$\cite{kang2020quantitativeGenetics,fisher1918quantitative}.  
While \emph{blending inheritance} theory, where parents traits are mixed to form the offspring features, is long considered obsolete when discussing individual genes and alleles
~\cite{dobzhansky1951genetics}(p.52),
quantitative genetics studies the many quantitative phenotypic traits which may manifest averaging effects. Such traits include lifespan, weight, strength, speed, and many others. Quantitative traits are affected by vast genetic information, commonly represented in Quantitative Trait Loci (QTL) which may include multiple genes, regulatory sequences, epigenetic effects, etc. 
The phenotype polygenic aggregation involves, among others, operations that resemble adding and averaging the various genetic factors within the organism. Furthermore, the way offspring QTL information is determined from the merging of parents' QTL is known to be highly complex~\cite{cui2004mappingQTL,niu2022inheritanceQTL}, and cannot  yet be reduced to a small set of rules; the model makes a modest attempt to reflect this complexity. 

Throughout this paper, we use the term \emph{an encoding} to refer to an object or structure that encodes, or represents, another object or concept. Thus, an encoding is synonymous here with the term \emph{a code}, which we presently reserve for the broader concept of \emph{species interaction code}. The use of the term "an encoding" as an instance, distinguishes it from the concept of ``a code'' as is used by Barbieri and others, where it refers to the entire mapping between domains~\cite{barbieri2015codeBiologyBook}. 

While the model does not aim to mimic meiosis or gamete formation, we opted to include in the model an elaborate step of recreating a parent's encoding that will then serve as a seed for subsequent reproduction steps, rather than sufficing with straightforward copying of the parent’s encoding. 

\subsection{Information aggregation by set operations} 

While the relation to quantitative phenotype traits allows some measure of numerical calculations on traits, we wanted the determination of offspring trait encodings to be carried out largely by set operations (like set union) on parent encodings, relying less on  pure numerical operations on the two sources. 

The present setting is amenable to extensions using more elaborate methods for deriving  child traits from parents' traits. 

Note that pure averaging of a single trait
represented by a random variable 
whose copying is subject to error with some probability, 
would be analogous to combining two one-dimensional random walks, with the same step size and equal probabilities for stepping in either direction. 
When two such random walks, starting at different origins, are combined by stepping in parallel and averaging the location, the resulting mean position is the average of the two origins. More importantly, the position variance is half the (equal) variance of the two original random walks. This is aligned with the known fact that bi-parental reproduction reduces the variance within the offspring generation.

\subsection{Minimalism, abstraction and accessible language}

The model is not intended to be a full model of what happens in nature.  In fact, we try to show that in situations of noisy inheritance, it is the very merging of parent encodings  that enables the cross-generation conservation of ancestral traits.
Examples of elements of the abstraction that we employed include: (i) 
The trait of interest and its encoding both involve real numbers. Extending the arguments to pure symbolic entities, or concrete objects like molecules, is left for a future research; (ii) 
The inevitable variation in natural reproduction is modeled by replacing direct copying with sampling from a normal distribution around the source value; (iii) the bi-parental reproduction is unisexual, i.e., there are no distinct mating types.

Since here we aim to only show relations between certain values, rather than establishing those values, we adopt the ``story proof'' argumentation style, as defined and used in~\cite{blitzsteinHwang2015probability}; this style relies on text of intuitive, yet well founded, logic statements to support the claims, and keeps mathematical formulas to a minimum. In addition, the occasional use of terms that are commonly applied to populations of living organisms is only for convenience.

In an attempt to align with natural processes, in the mathematical analysis and the simulation examples below, the summing and averaging of encoded traits is not carried out by of the very process of of fusing parental genetic material (syngamy, in nature), but by the organism development process that interprets the combined codes. Furthermore, the encoding that supports this process works well also for mono-parental reproduction.

\section{Problem Formalization}\label{sec:math} 
In this section we define the bi-parental and mono-parental reproduction methods in the context of a specific model.

\subsection{The formal model}

Consider the following synthetic model of a population of entities, also referred to as individuals. 
The size of the population is kept fixed at a number $n$. 
We label the entities in a cohort as 
$\{e_1,e_2,...,e_n\}$. 
The analysis does not include considerations of time. 
The population reproduces in methods discussed in detail below; most basically, in every reproduction step,  
a set of $n$ entities, referred to as the cohort of the next, or child, generation, appears instantly and synchronously. 

In the present discussion it does not matter whether the parent generation persists or disappears as long as the entities in the parent generation do not reproduce after the appearance of the child generation. 
For simplicity, we assume that the entire cohort of each parent generation  disappears immediately and synchronously upon the appearance of the cohort of the child generation. 
One reason for this choice is to avoid dealing with inter-generation mating in a population in which diversity among co-existing distant generations is unbounded.

All entities are associated with exactly one ``phenotype'' quantitative trait $P$, which is manifested as a real number. Different entities may have the same or different  values for $P$.
For illustration, the abstract entities may be thought of as fixed-width sticks where $P$ is their length, or as lumps of material where $P$ is their weight. Let $p$ denote any value of $P$, and, for example, $p_i$ is the value of $P$ in the entity $e_i$. 
With each new generation, under the two reproduction methods, the $p_i$ values and their distribution within each new cohort may change.
In each cohort $c$ we are first interested in the mean $p_c$ 
of the $p_i$ values within the cohort:
$$p_c = \frac{1}{n}\sum_{i=1}^{n}p_i~~.$$ 

In each parent generation $c$, one can view $p_c$ as representing a common trait, or capability, which is manifested with some diversity in the living individuals. 
(If the value of $p$ in an individual is essential for survival within the species' ecosystem, then the distribution of the trait $P$ among the members of the next generation $c'$ affects the likelihood that individuals from $c'$ will be able to participate in ecological networks in which their parents from $c$ no longer function. Note however, that such survival considerations are not part of the model, analysis, and simulations.)

The ``genotype'' encoding of $P$ in each individual $e_i$
is a set $g_i$ of $k$ real values which are also referred to as the points in $g_i$.  We allow the rare case of any two points in a set being equal to each other, without resorting to other mathematical terms like $g_i$ being a multiset or a tuple. The integer $k$ is a fixed parameter in the analytical model, and it is large, say, $k \ge 100$; we also constrain $k$ to be an even number. 

The phenotype manifestation of the trait $P$ in $e_i$ is the mean of the points in $g_i$. 
That is, given an individual $e_i$, $1 \le i \le n$, with genotype encoding $g_i=\{x_{i_1},x_{i_2},...,x_{i_k}\}$,its phenotype value $p_i$ is computed as follows: 
$$p_i = \frac{1}{k}\sum_{\ell=1}^{k}x_{i_\ell}~~. $$ 
\subsection{Reproduction methods}

We model and compare the effects of two different reproduction processes, termed \emph{Mono-parental Reproduction}~(MR, hereafter) and 
\emph{Bi-parental Reproduction}~(BR, hereafter). 

Below we describe one reproduction step of an entire generation, termed \emph{a generational transition}, in each of the methods.

\vspace{0.8cm}
\xn \textbf{Mono-parental Reproduction (MR)}

\begin{enumerate}
\item{The input is a set $c$ of the current population; create an empty set $c'$.}
\item{Repeat $n$ times to create $n$ children:}
\begin{enumerate}
\item{With equal probability, pick a single individual $e_i\in c$, by picking a random integer 
$1 \le i \le n$; the selected individual will be the single parent of the child created in this iteration; in separate iterations, the same individual $e_i$ may be selected again. Note:  hereafter, when the distribution of a random choice is not stated, it is assumed to be uniform, with equal probability.}
\item{Let $g_i=\{x_{i_1},x_{i_2},...,x_{i_k}\}$ be the genotype encoding of $e_i$. Compute the mean of  $g_i$, which is equal to the phenotype of $e_i$:  
$$p_i = \frac{1}{k}\sum_{\ell=1}^{k}x_{i_\ell}~~.  $$ }
\item{Compute a new set $g'$ by sampling $k$ real numbers, 
$g'=\{x'_1,x'_2,...,x'_k\}$, 
subject to normal distribution with mean $p_i$ and \stdev $h$ (and variance $h^2$) 
for some fixed global parameter $h$ which specifies and constrains the noise, or error, that may be introduced in the reproduction processes.}
\item{Create an entity $e'$ whose genotype encoding is $g'$, and its phenotype is the mean of the points of $g'$, and add it to $c'$.} 
\end{enumerate}
\item{After completing $n$ iterations: discard the parent set $c$; rename the child set $c'$ to $c$; discard $c'$.}  

\end{enumerate}

\vspace{0.2cm}
\xn \textbf{Bi-parental Reproduction (BR)} 

\begin{enumerate}
\item{As in MR, the input is a set $c$ of the current population; create an empty set $c'$.}
\item{Repeat $n$ times to create $n$ children:}
\begin{enumerate}
\item{Pick a  random pair of two separate individuals, $e_u, e_v \in c,~u \neq v, 1 \leq u,v \leq n$  out of the $\frac{n(n-1)}{2}$ possible pairs in $c$. 
The pair $e_u, e_v$ will be the joint parents of the child created in this iteration; in separate iterations these two individuals may participate in other pairs or again in the very same pair.} 
\item{Let the respective genotype encodings of $e_u$ and $e_v$   be \\
$g_u=\{x_{u_1},x_{u_2},...,x_{u_k}\}$ and  \\
$g_v=\{x_{v_1},x_{v_2},...,x_{v_k}\}$, and let $p_u$ and $p_v$ be the respective means of these genotype encodings; again, these mean values are  conveniently equal to the respective phenotypes.
}
\item{Create two interim (``gamete'') encodings $g'_u$ and $g'_v$ by randomly selecting $\fkt$ points, under normal distribution with mean $p_u$ and \var $h^2$, and $\fkt$ points under normal distribution with mean $p_v$ and \var $h^2$, respectively; let $g'$ be the union of these two sets, $g'=g'_u\cup g'_v$.
We term this process \emph{merging} of the parents'
encodings. Note  that this merging involves a sampling step, and does not rely on copying of elements of $g_u$ or $g_v$ into $g'$.}
\item{Create an entity $e'$ whose genotype encoding is $g'$, and its phenotype is the mean $p'$ of the points of $g'$, and add it to $c'$.}  
\end{enumerate}
\item{As in MR, after completing $n$ iterations: discard the parent set $c$; rename the child set $c'$ to $c$; discard $c'$.}  
\end{enumerate}

As stated earlier, neither MR nor BR employ considerations of Natural Selection and classical fitness metrics like an individual's probability of reproduction and its expected number of offspring over its life time, and BR does not make use of mating types or  mate selection.

\subsection{Model realization example}
Below we illustrate the above formal model with an example realization in an imaginary species of intelligent and capable animals. 
One of the traits that sustains this species is their ability to build every year a mud hill of ``exactly'' a certain height, as traps for some creature that they feed upon. The way they remember the right height is that each mature individual carries with them a bunch of sticks that they got from their parents. The individual builds each mud hill so that its height is close to the mean length of its set of sticks. Old mud hills from the previous year are not available for comparison, as they have been washed away by the rains. When young individuals mature, they leave their parents to look for new territory. Before they leave, the parents (one or two of them, as the case may be) equip the young with a new bunch of sticks, cut as best as they can to the mean length of the parents' bunch.  In the case of two parents, each one gives the child half the number of sticks based on that parent’s set. The parents cannot give their own sticks to the  offspring,  as they have to keep their own sticks for several more years. The individuals are solitary most of the time, so each one needs their own set of sticks. 
Individuals of the species use sets of sticks  for this purpose, rather than just one, as  the sticks may break or get lost.  In addition, the parent’s skill and the available material limit the precision of each cut stick; having several sticks, some too short and some too long, is a  way for conveying the height of the mud hill that is  desired by  the individual preparing the new set.

\subsection{A numerical example}
For illustration, below we go through the core elements of MR and BR  using a small-scale example. 
\begin{enumerate}
\item \textbf{Parameter Setting:}
Assume that in the parents' cohort $c$ ,  $P$ is normally distributed 
in the parent generation
with mean  
$p_c=2.0$, and with \stdev $\sigma_c = 0.2$ 
The number of points in a genotype is $k=4$; the points in a genotype set are shown here in a sorted order for easier reading.
The standard deviation of the normal distribution of the reproduction noise is $h=0.15$; 
\item{
Let individual \textcolor{blue}{$e_1$} have a genotype encoding of  
\textcolor{blue}{$g_1=\{1.70,1.80,1.90,2.10\}$} whose mean and phenotype is \textcolor{blue}{$p_1=1.875$}; }
\item{\textbf{MR of \textcolor{blue}{$e_1$}:} Compute the mean of the genotype, yielding again \textcolor{blue}{$p_1=1.875$}. Draw a sample of $k=4$ points around this value as a mean with the above \stdev $h=0.15$, yielding say, 
\textcolor{blue}{$g'=\{1.61,1.71.1.91,2.01\}$}, whose mean (and phenotype of child \textcolor{blue}{$e_{\text{1,MR}}'$}) is  \textcolor{blue}
{p'=1.81}};
\item{Let individual \textcolor{red}{$e_2$} in the parent cohort $c$ have genotype encoding of 
\textcolor{red}{$g_2~=\{1.95,2.15,2.25,2.35\}$}, whose 
mean and phenotype is  \textcolor{red}{$p_2=2.175$}. 
}

\item{\textbf{BR of \textcolor{blue}{$e_1$} and \textcolor{red}{$e_2$}:} 
We sample $\fkt = 2$ points around the mean \textcolor{blue}{$p_1$}, and $2$ points around the mean \textcolor{red}{$p_2$}, both with the noise \stdev $h$, yielding, say, \textcolor{blue}{$g'_1=\{1.8,2.1\}$}, and 
\textcolor{red}{$g'_2=\{2.02,2.18\}$}, respectively. The union of these two samples  yields $g'=\{\textcolor{blue}{1.8},\textcolor{red}{2.02},\textcolor{blue}{2.1},\textcolor{red}{2.18}\}$}, whose mean (and phenotype of the joint offspring $e_{\text{1,2,BR}}'$) is $p'=2.025$.

\end{enumerate}
 
\section{Comparing the Two Reproduction Methods}\label{sec:comparison}
In comparing  MR and BR in a particular generational transition from a parent cohort $c$ to a child cohort $c'$, we are initially interested in a basic metric $d$ defined as the distance between the phenotype trait of a random individual in $c'$, denoted $p'_i$, and the common trait $p_c$ of the parents' cohort: $$d_i = |p'_i - p_c|~ .$$ 

For a given parent cohort $c$, let  $d_\mr$ and $d_\br$ be the expected values $E[d_i]$ under the MR and BR processes, respectively, when considering all possible 
$c \shortrightarrow c'$ generational transitions that could emanate from $c$ in one reproduction step, and random choices of $e'_i$ individuals within $c'$.

For a given parent cohort $c$, let  $\sigma^2_\mr$ and $\sigma^2_\br$ be the expected values $E[\sigma^2_{c'}]$, the \var of $P$ within $c'$ under the MR and BR processes, respectively, when considering all possible $c \shortrightarrow c'$ generational transitions that could emanate from $c$ in one reproduction step under the  respective reproduction process.

\vspace{0.4cm}
\xn\textbf{Proposition 1.} 
Let $c$ be a cohort of $n$ entities in the above model;  assume that $n$ is large and 
that the distribution of the $p_i$ values of individuals in $c$ approximates a random sample from a normal distribution with mean $p_c$ and some \var $\sigma^2_{c}$; 

\xn Then, $d_\br < d_\mr$.

\xv

\xv 

\vspace{0.4cm}
\xn\textbf{Proof.}

\xv

\xv 

\xn \textbf{A. Compute $d_\mr$.}

\xv 

Let $c'_\mr$ be a random child cohort 
of~$c$, selected from all possible MR generational transitions  with  
$c$ as a parent.  

\xv

Let $e'_{i_\mr}$ be a random individual in $c'_\mr$.  

\xv 

Let $g'_{i_\mr}$ be the genotype encoding of $e'_{i_\mr}$, and  let $p'_{i_\mr}$ be its phenotype.  

\xv

Let $e_{i}$ be the element of $c$ 
that served as the single parent of $e'_{i_\mr}$ in this MR generational transition. 

\xv 
Let $p_i$ be the phenotype of $e_i$ (it is also the mean of the genotype encoding $g_i$ of $e_i$). 

\xv 

The genotype encodings $g'_{i_\mr}$ of all possible direct MR children of $e_i$, can be considered as samples of size $k$ from a normal distribution with mean $p_i$ and \var $h^2$,  where $h$ is the noise parameter defined above. 

According to the central limit theorem~\cite{blitzsteinHwang2015probability}(p.435), the means of these samples, namely $p'_{i_\mr}$, are distributed normally with mean $p_i$ and \stdev $\frac{h}{\sqrt{k}}$ (and \var $\frac{h^2}{k}$). 

\xv 

We are facing now a compound, nested distribution, that involves (i) the random selection of an 
 $e_i$, which is close enough for our purposes to drawing a random value
  $p_i$ from the original normal distribution of $c$ with mean $p_c$ and \var $\sigma^2_c$, and (ii)  based on the selected $p_i$ and the noise parameter, randomly selecting the sample $g'$ of size $k$ and computing its mean $p'_{i_\mr}$.

We can  now examine the distribution of the random variable $p'_{i_\mr}$ as drawn from the original distribution of $c$. 
According to the law of total variance (a.k.a. Eve's law~\cite{blitzsteinHwang2015probability}(p.401)), this is a normal distribution with mean $p_c$  and \var that is the sum of the variances, hence 
$$\sigma^2_{c'_\mr} = \sigma^2_c + \frac{h^2}{k}$$

In general, the average absolute deviation of a random sample from the mean of a normal distribution with \stdev $\sigma$ is 
$\sigma \cdot \sqrt{\frac{2}{\pi}}~~$~\cite{geary1935deviation}.
Therefore, the expected value of the absolute distance of $p'_{i_\mr}$ from $p_c$ is 

$$d_\mr = E[~|p'_{i_\mr}-p_c|~] = \left(\sqrt{\sigma^2_c + \frac{h^2}{k}}\right)\cdot \sqrt{\frac{2}{\pi}}~~~~~.$$

\xn \textbf{B. Compute $d_\br$.}

\xv 

We now examine the expected effects of all possible BR generational transitions with  
$c$ as a parent cohort.   

\xv

Let $c'_\br$ be a random child cohort 
of~$c$ within all possible BR generational transitions.  

\xv

Let $e'_{j_\br}$ be a random individual in $c'_\br$.  

\xv 

Let $g'_{j_\br}$ be the genotype encoding of $e'_{j_\br}$ and and let $p'_{j_\br}$ be its phenotype.  

\xv

Let $e_u$ and $e_v$ be the two elements of $c$ 
that served as the parents of $e'_{j_\br}$ in this BR generational transition, and let  $p_u$  and $p_v$ be the means of their genotype encodings, respectively.

\xv 

The genotype encoding $g'_{j_\br}$ was created as a union of two random samples $g'_u$ and $g'_v$, each of size $\fkt$,  around the means $p_u$  and $p_v$, respectively. 
Let $p'_u$ and $p'_v$ be the means of  $g'_u$ and $g'_v$, respectively. 

The phenotype $p'_{j_\br}$ is the mean of the union of  $g'_u$ and $g'_v$, and is thus the average of the means of these two equal size samples: $$p'_{j_\br} = \frac{p'_u  + p'_v}{2}$$

Let us now analyze the random selection of $p'_{j_\br}$ compounding the randomness inherent in the way  $g'_u$ and $g'_v$ were sampled, and the nested randomness of $p_u$ and $p_v$ as selected from the original distribution of $c$. 

The distribution from which both $p_u$ and $p_v$ were sampled is normal with mean $p_c$ and \var $\sigma^2_c$.

The sets $g'_u$ and $g'_v$  were drawn from distributions with means $p_u$ and $p_v$ respectively, both with \var $h^2$. 
According to the central limit theorem  again, the distributions of $p'_u$ and $p'_v$ are normal with mean $p_u$ and $p_v$ respectively, and \stdev $\frac{h}{\sqrt{\fkt}}$ (and \var $\frac{h^2}{(\fkt)}$).

Compounding each of these two distributions  separately with the underlying distribution of $c$ we get that $p'_u$ and $p'_v$ are distributed normally with mean $p_c$ and \var that is the sum $\sigma^2_c + \frac{h^2}{\fkt}$~. 

Clearly the mean of the random variable   $p'_{j_\br}$ is $p_c$ 
since it is \emph{half} of the sum
$E[p'_u]$ and $E[p'_v]$ (each of which is equal to $p_c$),  and its \var is a quarter of the variance of their sum, as it is the square of the scaling by half of the variable and the mean~\cite{blitzsteinHwang2015probability}(p.159):   
$$\sigma^2_\br = \frac{1}{4}\cdot 2(\sigma^2_c + \frac{h^2}{\fkt}) = \frac{\sigma^2_c}{2} + \frac{h^2}{k}~.$$

\xn As before, the average absolute deviation is given by 

$$d_\br = E[|p'_{j_\br}-p_c|] = \left(\sqrt{\frac{\sigma^2_c}{2} + \frac{h^2}{k}}\right)\cdot \sqrt{\frac{2}{\pi}}~~.$$

\xn \textbf{C. Comparing $d_\br$ and $d_\mr$.}

\xv 
The expressions for $d_\mr$ and $d_\br$ are the same except for the first term in  $d_\br$ being 
$\frac{\sigma^2_c}{2}$ where in $d_\mr$  it is $\sigma^2_c$. Hence,  $$d_\br < d_\mr~. ~~~~~~~~~~~~~\square$$. 

\xv  

Note that an intermediate result in the above calculations is that $\sigma^2_\br < \sigma^2_\mr$. This aligns well with the established fact, as was discussed earlier, that in nature, sexual reproduction yields a lower variance \emph{within} the new generation than does mono-parental reproduction. In the present context, such smaller values of $\sigma_{c'}$  help individuals in $c'$ step in and replace \emph{each other} when needed, independently of the magnitude of their differences from their parents  and from earlier generations.

\section{Visual Illustration by Simulation}\label{sec:simulations}
Below we present the results of a few simulation run examples. These simulations are meant to serve as an accessible visual illustration for the general effects predicted by the above mathematical analysis; they were not subjected to elaborate quantitative analysis. For each simulation case we  show here just one run; multiple runs using the same parameters and differing only by the pseudo-random numbers used at various steps in the reproduction process, yielded similar results.  

 \subsection{Reproducing a single real number via a set of points}

\begin{figure}[ht] 
	\centering
	\includegraphics[width=0.95\linewidth]{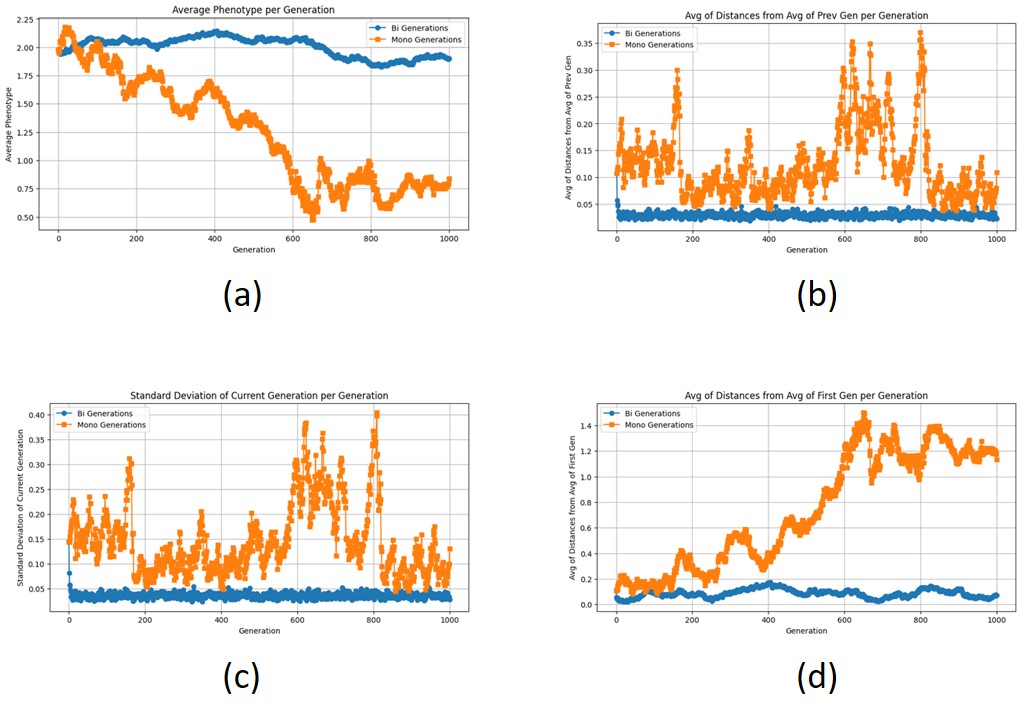}
	\caption{Simulating points reproduction. See explanation in text.}
 \label{fig:pointsGraph}
\end{figure}

 The first set of simulations follows directly the model described in 
Section~\ref{sec:math} demonstrating the reproduction of a real number through an intermediate encoding by a cohort of points.
The program source is available in~\cite{bipar2024colab}.
Figure~\ref{fig:pointsGraph} depicts the results. 
The orange color graphs refer to mono-parental reproduction (MR), and the blue to bi-parental reproduction (BR). 
The simulation was  run with the following parameters.
\begin{itemize}
    \item Cohort size: $n=32$ 
    \item Encoding set size: $k=30$
    \item Number of generations: $1000$ 
    \item The Initial population, i.e., the cohort of generation $1$, was created by random sampling of points from a normal distribution with mean $2.0$ and standard deviation of $sd_c=0.2$.  
    \item Initial common phenotype is thus: $\sim2.00$.
    \item Reproduction error was produced by random sampling from a normal distribution where the mean is the parent's phenotype, and the standard deviation is $h=0.15$.
\end{itemize}

We observe the following: (a)~The average of the phenotypes of the entire cohort is closer to the original phenotype of approximately $2.00$ under BR than  under MR. 
(b)~On average, the phenotypes of individuals of each generation, are closer to the average of the immediately preceding parent generation in BR than in MR.
(c)~The standard deviation within each generation cohort is smaller under BR than under MR; i.e., the similarity of members of a given generation to each other is higher under BR than under MR.  (d)~On average, under BR, the phenotypes of individuals of each generation, are closer to the average of the original ancestral generation  than under MR. 

\subsection{Reproducing images of printed text}

The second set of simulations is inspired by the way the survival of ancient manuscripts over hundreds and thousands of years was enabled by reproduction, i.e., copying. Both ``mono-parental'' and ``bi-parental'' techniques were employed by scribes: most often copying from a single source, but sometimes collating multiple sources
~\cite{skeat1999codexSinaiticusCollation}(p.585;617). (Note: we distinguish here between collation of several sources in the process of copying text, from the collation of sources by history researchers reconstructing an original document.)
In this example we applied the reproduction techniques described above for the reproduction of images of text containing one word. 
We introduced a rudimentary measure of the preservation of traits as the success rate (across the cohort)  of  image recognition software (gpt-4-turbo model by OpenAI) in recognizing the text. This metric also hints at the importance of traits in an organism's interactions; in the present case, the interaction can be seen as the sending and receiving of messages.  

\xv 
\xn The details of the process are as follows.  Consider first the following two functions:

\xv 
\xn \textbf{Function F1. Noisy encoding of an image.} Given an input image, create a copy of it; the copy may be of smaller, same or larger resolution; This step can be seen as a representation of the action of a person wishing to preserve a precious image or text by hand-copying it.
Then, add noise to the copy by sampling random values from a normal distribution with mean 0 and standard deviation $h$, and adding them respectively to each of the three RGB values of all pixels of the copy; round the result to integer and clip the values at 0 and 255. This step may represent the imperfection involved with physical copying; image resizing also adds noise, but since the process is deterministic, 
it is identical in in all copies, and hence is less significant for our purposes.

\xv 
\xn 
\textbf{Function F2. Noisy computation of a ``phenotype''.} Given $k$ copies of an image, as may be created using F1 above, compute the average of all respective pixel values; 
add noise to the result in the same manner as in F1, and resize the resulting image to the original resolution. This step may represent the collation process done by a person creating a fresh copy based on several sources, where each may have mistakes and defects. 

\xv 

\xn The parameters used in the simulation are: seed image 
 is an image of the word ``TEACHING'', typed in the font Calibri Light, black and white only, at 
resolution 100x100 pixels; copies generated by F1 are 400x400 pixels; noise standard deviation is $h=20$; cohort size is $n=8$; and, the number of copies in the ``image genotype'' encoding is $k=8$. 
 The simulation steps are as follows:

\begin{figure}[t] 
\hrule 
	\centering
	\includegraphics[width=0.8\linewidth]{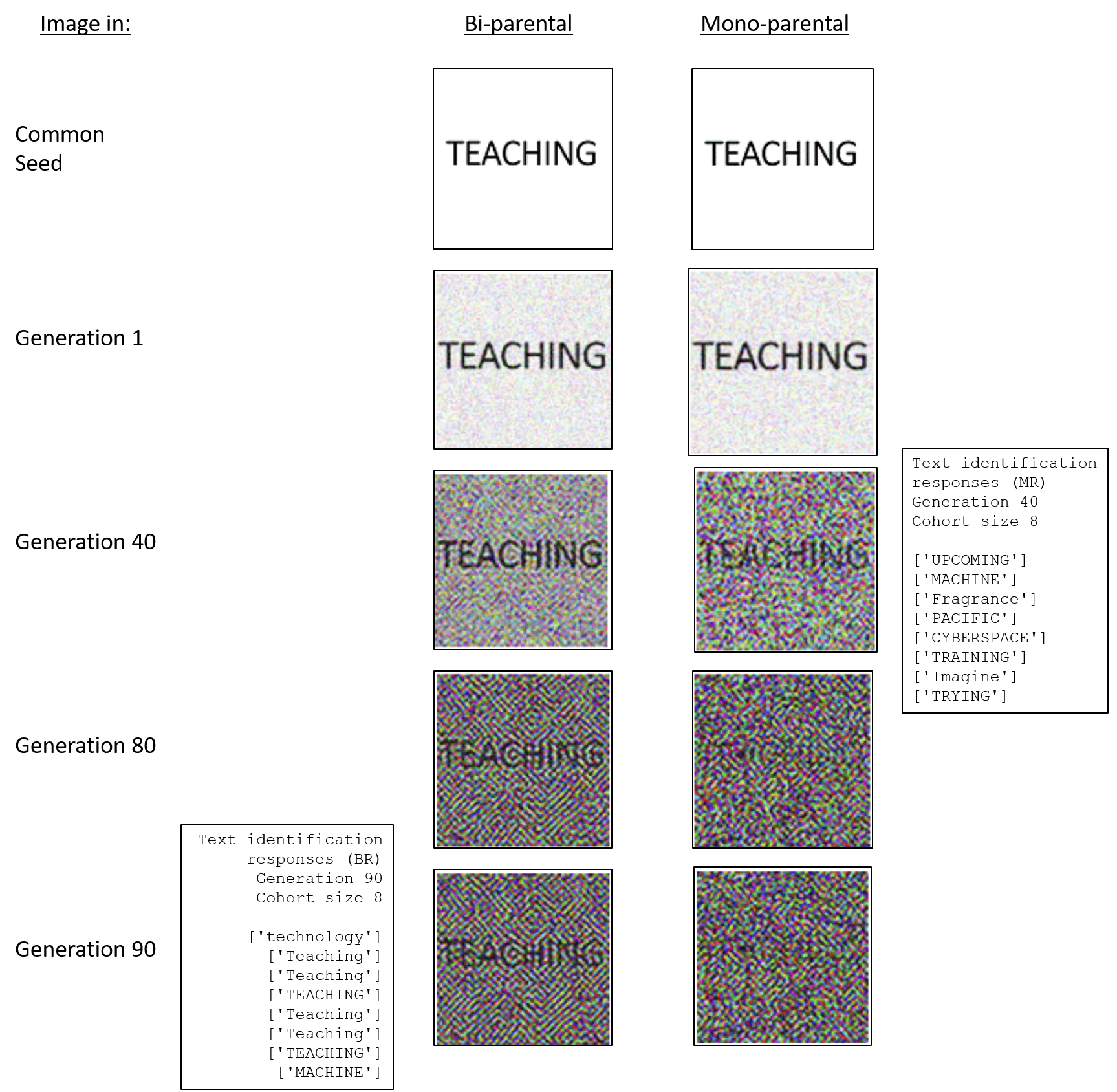} \\ ~~ \xv 
	\caption{Simulating reproduction of an image of printed text. See explanation in the body of the article.}
\hrule 
 \label{fig:teachingZeroTo90}
\end{figure}

\xv 

\xn \begin{enumerate}
\item{Start with a seed image.}
\item{[Creating base cohort.]  Repeat the following $n = 8$ times:} 
\begin{enumerate}
\item{[Compute ``child genotype''.] Create $k$ copies, applying F1 above $k$ times to the seed image.} 
\item{[Compute ``child phenotype''.] Create one child image by applying F2 to the above image genotype.} 
\end{enumerate}
\item{Copy the above base cohort to create  generation 1 of the BR process and generation 1 of the MR process.} 
\item{[BR process.] Repeat the following for 90 generations}
\begin{enumerate}
    \item{[ Create next generation.] Repeat $n = 8$ times:}
    \begin{enumerate}
    \item{Randomly select two individuals from the current generation.}
    \item{Apply F1 to each of these two individuals $\frac{k}{2}=4$ times.}
    \item{Apply F2 to the $k=8$ copies, yielding a new child individual.}
    \end{enumerate}
    \item{Every 10th generation run image recognition on each of the cohort’s $n=8$ images, using OpenAI API with the model gpt-4-turbo and the prompt: “Your role is to identify the word in the image. Please provide a 1-word answer.”}
\end{enumerate}
\item{[MR process.] Repeat the following for 90 generations}
\begin{enumerate}
    \item{[Create next generation.] Repeat $n=8$ times:}
    \begin{enumerate}
    \item{Randomly select one individual from the current generation.}
    \item{Apply F1 to this individual $k=8$ times.}
    \item{Apply F2 to the $k=8$ copies, yielding a new child individual.}
    \end{enumerate}
    \item{Every 10th generation run the same image recognition task on the $n=8$ cohort's individuals as under BR.}
\end{enumerate}
\end{enumerate}

Figure~\ref{fig:teachingZeroTo90} depicts a sample of the resulting images, with one sample image from each sample generation.

Under  BR the LLM identified the word ``TEACHING'' throughout more than 70 generations, while under MR it failed the identification already on or before generation 40. In the 90th generation, under BR, the LLM did not identify the correct word but still identified a word (e.g., ``TECHNOLOGY'' instead of ``TEACHING''), while under MR it did not identify that the image contained text and responded that ''The image appears to be a stereogram  or a pattern designed to create a visual illusion making it difficult to determine if there’s any specific word or object within it.  If there's a particular way I should analyze this please let me know!''.)

\begin{figure}[ht] 
\hrule 
	\centering
	\includegraphics[width=1\linewidth]{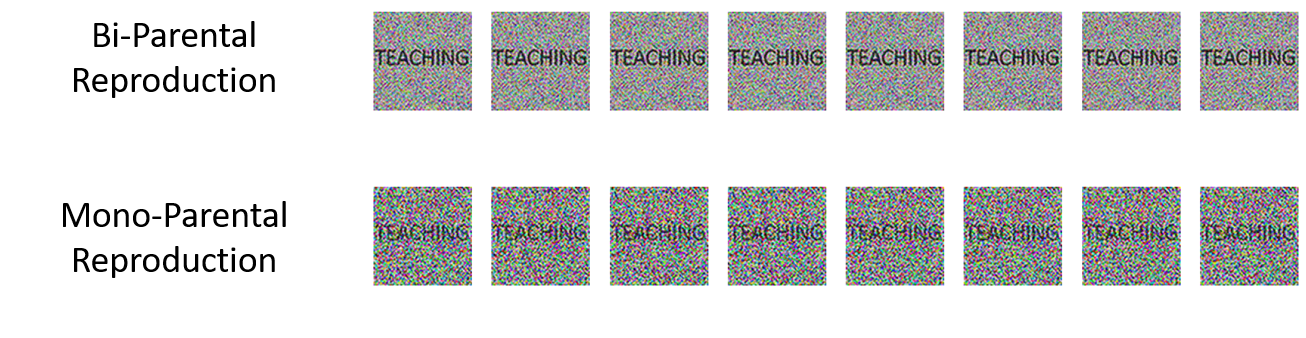}
	\caption{The 8 individuals of generation 40 under BR and MR. See discussion in the body of the article about retention of common traits, both ancestral and emergent.}
\hrule 
 \label{fig:cohortCommonTraits}
\end{figure}

We have also looked casually for effects regarding additional common traits across the various cohorts. We noticed that traits like contrast between the letters and their background, or 
the horizontal line of the letter T were preserved better under BR than under MR (see Figure~\ref{fig:cohortCommonTraits}).
Under BR, we have also noticed conspicuous and sustained emergent traits, like a circle shape inside the letter C or a third vertical line in the letter H, where under MR we could not readily find any such emergent traits.

\subsection{Reproducing images of flowers} 

\begin{figure*}[t] 
\hrule
	\centering
	\includegraphics[width=1\linewidth]{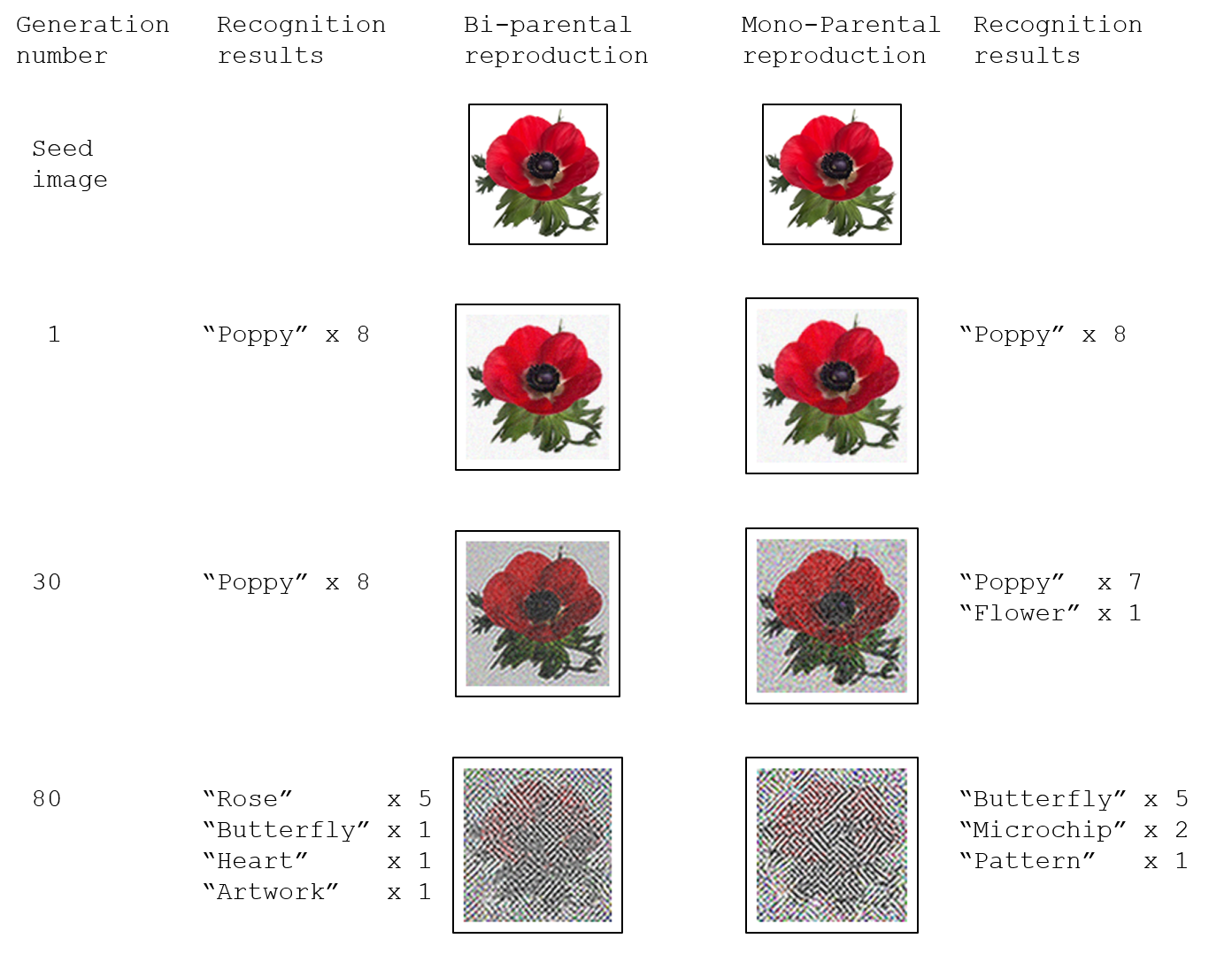}
	\caption{Simulating reproduction of an image of a flower. See explanation in text.}
\hrule
 \label{fig:poppygens}
\end{figure*}

In this illustration example we were inspired by object recognition in nature, as is done, for example, by insects species that pollinate only certain plant species. Figure~\ref{fig:poppygens} depicts  reproducing an image of a flower under BR and under MR.  The simulation uses the same process as in the case of text image, with the following parameters:
The starting image was one labeled as  anemone flower, downloaded from 
{(\small \url{https://it.pinterest.com/pin/red-anemones--797207571571931709/}}),
and used under fair use license; the image was then cropped and resized to 100 x 100 pixels; noise standard variation is $h=10$; cohort size is $n=8$; encoding set size is $k=8$; number of generations is $80$; image recognition prompt is: “Your role is to identify the object in the image. Please provide a 1-word answer.".

One can see that, under BR, images were assigned the same label as the first one, ``Poppy'', for more generations than under MR. (We ignore the fact that the LLM's first identification did match the label ``anemone'' originally assigned to the image.) 
We also observe that certain core traits, like being a flower were better preserved under BR. 

Note: The images shown here are illustrative examples. 
While in all runs the image traits were sustained for longer under BR than under MR, some intermediate generations showed a local advantage for MR. Such results are to be expected, first, due to the random nature of the inserted noise, and perhaps more so, due to the possibility that in a particular cohort, an individual may reproduce several times, dominating the traits in the next generation.

\section{Conclusion and Next Steps}

We have shown that merging and averaging effects in bi-parental, or sexual, reproduction can contribute to conservation, in the offspring generation, of traits that are common in the parent generation, in a way that complements known biological processes like reproductive limitations and DNA repair. Confirming the observation in nature, and investigating it in more elaborate models, including multiple traits, ecosystem interactions, and incorporating more biological details, remains as future work.  

\section*{Acknowledgements}
We thank Netzer Moriya and Hagai Cohen for valuable discussions. 
This research was funded in part by an NSFC-ISF grant to DH,  issued jointly by the  National Natural Science Foundation of China (NSFC) and the Israel Science Foundation (ISF grant 3698/21). Additional support was provided by a research grant to DH from Louis J. Lavigne and Nancy Rothman, the Carter Chapman Shreve Family Foundation, Dr. and Mrs. Donald Rivin, and the Estate of Smigel Trust.



\section*{Author contributions} 
A.M. initiated, designed and led this work, carried out the mathematical analysis, developed a preliminary version of the simulation software, analyzed the results, and wrote the manuscript;
S.S. contributed to the design and analysis of the simulation experiments, and developed the software; 
I.R.C. conceived ideas underlying this research direction; 
I.R.C. and A.M. developed the encoding-based research focus;
D.H. contributed to the underlying research and the drafting of the manuscript;
the research was conducted and funded within the group of D.H.;
all authors discussed the results and commented on the manuscript.

\end{document}